# Online Critical-State Detection of Sepsis Among ICU Patients using Jensen-Shannon Divergence

Jeffrey Smith[1], Christopher Josef MD[2], Yao Xie, PhD[1], Rishikesan Kamaleswaran, PhD[1,2]
[1]Georgia Institute of Technology, Atlanta, Georgia, USA;
[2]Emory University School of Medicine Atlanta, Georgia, USA

**Abstract**

*Sepsis is a severe medical condition caused by a dysregulated host response to infection that has a high incidence and mortality rate. Even with such a high-level occurrence rate, the detection and diagnosis of sepsis continues to pose a challenge. There is a crucial need to accurately forecast the onset of sepsis promptly while also identifying the specific physiologic anomalies that contribute to this prediction in an interpretable fashion. This study proposes a novel approach to quantitatively measure the difference between patients and a reference group using non-parametric probability distribution estimates and highlight when abnormalities emerge using a Jensen-Shannon divergence-based single sample analysis approach. We show that we can quantitatively distinguish between these two groups and offer a measurement of divergence in real time while simultaneously identifying specific physiologic factors contributing to patient outcomes. We demonstrate our approach on a real-world dataset of patients admitted to Atlanta, Georgia's Grady Hospital.*

**Introduction**

Sepsis is a severe medical condition caused by a dysregulated host response to infection[1]. It has been estimated that worldwide, sepsis has affected 49 million individuals and was related to approximately 11 million potentially avoidable deaths, representing more than 20% of all global deaths[2]. In 2013, sepsis was the most expensive medical condition treated in the United States, accounting for almost $24 billion of all hospitalization expenses[3]. Even with such a high occurrence rate, the detection and diagnosis of sepsis continues to pose a challenge. The sepsis definition has been frequently updated over the years[1], yet its effective prediction serves as a difficult task where clinical outcomes still require adjudication from medical staff before treatment can commence. Among a survey of sepsis studies, Teng and Wilcox[4] remark that multiple criteria were employed to retrospectively categorize a patient as having the infection, indicating that the "gold standard" of today's definition[1] is still subject to interpretation. The union of these complexities in the concept of sepsis makes it even more difficult to recognize the condition immediately. Delays in the identification and initiation of treatment of sepsis have been shown to negatively impact patient outcomes. In a study from the New England Journal of Medicine, it was demonstrated that an hour increase in the time to administer treatment was linked to an increase in risk-adjusted in-hospital mortality (odds ratio, 1.04 hourly)[5]. It suffices to say that a clinician's ability to rapidly diagnose this infection may provide medical professionals additional time to intervene, thereby improving both patient outcomes and reducing the financial burden of healthcare.

Over the last decade, there has been a boom in the quantity of digital information contained in electronic health records (EHRs), which has allowed the use of healthcare data for activities such as anomaly detection. This technique, along with a variety of other machine learning and artificial intelligence technologies have grown more prevalent in the academic literature on sepsis during the past several years[6–9]. Numerous researchers have made significant contributions to this subject by utilizing a variety of machine learning approaches, frequently with models that performed well on their assigned tasks.

While machine learning approaches are efficient for diagnosis, they frequently vary in complexity, sometimes resulting in uninterpretable black box models. The interpretability of machine learning models' output is a key barrier to broader use of many algorithms in healthcare since medical practitioners must account for their medical choices. Thus, the challenge becomes one of accurately forecasting the onset of sepsis promptly while also identifying the specific physiologic anomalies that contribute to this prediction in an interpretable fashion. As a result, there is a critical need for interpretable models in the early diagnosis of sepsis[10].

Additionally, existing algorithms rely on population level abstractions for each instance of prediction tasks, therefore it may be difficult to understand precisely which features, dynamically over time, contribute to the onset, progression and ultimate resolution of this condition. Among methods that may be used to address this gap in knowledge are a class of methods known as anomaly detection which is the process of identifying abnormal or unexpected events behavior in datasets[11]. The purpose of this study is to compare septic and non-septic patients and to highlight when abnormalities emerge over a defined time period using a Jensen-Shannon divergence-based single sample analysis approach.

Research Objective: Our research objective is two-fold for this analysis:
- We seek to provide a quantitative measure of the divergence between an individual or subset of persons and a broader reference group in terms of real values
- We seek to examine the shift in probability distributions between patient trajectories over time to identify specific physiologic factors contributing to patient outcomes

**Contributions**

In this paper, we use Jensen-Shannon divergence across many layers of EHR data to detect health decline in real time for individual patients and patient subgroups and to give an interpretable, real-valued metric of this deterioration. We describe the problem using non-parametric probability distribution estimates for reference groups and individual patients, and then compute the divergence between these entities on a defined time basis. Our objective is to determine the key moment at which a patient is most likely to acquire sepsis while also identifying the factors that most influence this conclusion.

**Data**

The data were acquired retrospectively from 119,923 individual records from Grady Memorial Hospital, a public hospital in Atlanta, Georgia, for patients with a hospital admission between 2016 and 2020.

*Definition of Sepsis:*
Sepsis is defined as a life-threatening organ failure induced by a dysregulated host response to infection in accordance with the revised Sepsis 3.0 definition[1]. The retrospective diagnosis of sepsis was established by implementing the Sepsis-3 definition as described in our prior work[7]. Briefly, we implement the suspicion of infection criteria by identifying instances where the delivery of antibiotics in conjunction with orders for bacterial blood cultures occurred within a predetermined period. It is then determined that organ dysfunction has occurred when there is at least a two-point increase in the Sequential Organ Failure Assessment (SOFA) score during a specified period of time. The SOFA score is a numerical representation of the degradation of six organ systems (respiratory, coagulatory, liver, cardiovascular, renal, and neurologic)[12].

This study examined Grady Hospital patients admitted to the Intensive Care Unit (ICU) between 2016 and 2020. Patients were separated into two distinct cohorts based on a retrospective diagnosis of sepsis and the duration of each patient's stay in the critical care unit. The control cohort included patients who had a SOFA score of two or more at least 24 hours following admission to the ICU. While these individuals met the SOFA score requirements, they were not retrospectively diagnosed as having acquired sepsis. The secondary cohort, referred to as the treatment cohort, included patients with a SOFA score greater than two at least 24 hours after admission that were retrospectively identified as having acquired sepsis. In order to prevent problems with data imbalance in the analysis, a 48-hour time block was used for each patient. Hourly data for patients in the treatment cohort were stopped 24 hours after retrospectively being diagnosed with sepsis, whereas hourly data for patients in the control cohort were stopped 24 hours after receiving a SOFA score of two or higher. Patients under the age of 18 were not included in this analysis. The general characteristics of patients analyzed during this study are summarized based on cohort in Table 1.

**Table 1.** Baseline characteristics of study patients grouped by cohort

|  |  |  | Grouped by sepsis | |
|---|---|---|---|---|
| Variable |  | Overall | Control (Non-sepsis) Group | Treatment (Sepsis) Group |
| N |  | 893 | 749 | 144 |
| Age, mean (SD) |  | 54.8 (18.6) | 54.6 (19.1) | 55.7 (15.5) |
| Gender, n (%) | Female | 332 (37.2) | 290 (38.7) | 42 (29.2) |
|  | Male | 561 (62.8) | 459 (61.3) | 102 (70.8) |
| Race, n (%) | Asian | 10 (1.1) | 8 (1.1) | 2 (1.4) |
|  | Black or African American | 530 (59.4) | 456 (60.9) | 74 (51.4) |
|  | Hispanic | 44 (4.9) | 35 (4.7) | 9 (6.3) |
|  | White or Caucasian | 290 (32.2) | 234 (31.2) | 53 (36.8) |
|  | Other | 21 (2.4) | 15 (2.0) | 6 (4.2) |
| Weight (lbs), mean (SD) |  | 182.6 (48.7) | 182.2 (48.4) | 184.6 (50.3) |
| Length of stay (LOS): Hospital, mean (SD) |  | 15.1 (19.4) | 11.8 (9.1) | 32.7 (39.5) |
| LOS: ICU, mean (SD) |  | 6.3 (8.1) | 4.5 (3.6) | 15.5 (15.7) |
| No. days on ventilator, mean (SD) |  | 8.6 (12.9) | 2.4 (1.6) | 13.4 (15.5) |
| Mean SOFA score, mean (SD) |  | 1.2 (0.7) | 1.1 (0.6) | 1.7 (0.9) |
| Temperature, mean (SD) |  | 98.5 (0.9) | 98.4 (0.9) | 98.9 (1.1) |
| Pulse, mean (SD) |  | 88.1 (16.4) | 86.7 (16.2) | 95.0 (15.7) |
| Respiratory rate, mean (SD) |  | 18.7 (3.2) | 18.5 (3.0) | 20.0 (3.7) |
| Systolic BP, mean (SD) |  | 130.6 (15.7) | 130.4 (15.7) | 131.8 (15.6) |
| White blood cell count, mean (SD) |  | 10.6 (4.3) | 10.2 (3.5) | 12.7 (6.9) |
| Creatinine, mean (SD) |  | 0.9 (0.3) | 0.9 (0.3) | 0.9 (0.4) |
| Bilirubin TCount, mean (SD) |  | 0.8 (0.4) | 0.8 (0.3) | 0.9 (0.9) |
| Glasgow Coma Scale score, mean (SD) |  | 14.4 (0.9) | 14.6 (0.7) | 13.7 (1.3) |

**Methods**

We will define notation for the problem at hand. Individual patient encounters $p_k$ are comprised of several irregularly sampled features $x_{t_i,j}^{p_k}$ where $p_k$ for $k \in \{1,2,\ldots,K\}$ denotes an individual patient, $t_i$ for $i \in \{1,2,\ldots N_{p_k}\}$ represents each feature's recorded time following each patient's admission, and $x_j$ for $j \in \{1,2,\ldots M\}$ is the feature identifier. $N_p$ represents a patient's observation period length. These features include irregularly sampled vital signs and laboratory results. Vital signs in the ICU are normally recorded at an hourly interval and laboratory test are most commonly collected once every 24 hours, but often time vary in frequency based on the severity of the patient's illness and the medical judgement of the clinician. The final dynamic features used for this study are reported in Table 2.

**Table 2.** List of vital signs and lab features

| Vitals | Labs | |
|---|---|---|
| Glasgow Coma Score total | Alanine aminotransfer | Glucose |
| Oxygen saturation (SpO2) | Albumin | Hematocrit |
| Heart Rate | Alkaline phosphatase | Hemoglobin |
| Respiratory rate | Anion gap | International normalized ratio (inr) |
| Temperature | Aspartate aminotransferase | Magnesium |
| | Bicarbonate | Platelets |
| | Bilirubin | Potassium |
| | Blood urea nitrogen | Protein |
| | Calcium | Prothrombin time |
| | Chloride | Sodium |
| | Creatinine | White blood cell count |

Due to the asynchronous nature of these characteristics, observations are rarely made concurrently and can be represented as a sparse matrix of observations

$$\begin{pmatrix} x_{t_1,1}^{p_k} & \cdots & x_{t_1,M}^{p_k} \\ \vdots & \ddots & \vdots \\ x_{t_{N_{p_k}},1}^{p_k} & \cdots & x_{t_{N_{p_k}},M}^{p_k} \end{pmatrix}$$

where $M$ represents the total number of features for the encounter.

*Jensen-Shannon Divergence*
The Jensen-Shannon divergence (JSD), introduced by Burbea and Rao[13] and coined by Lin[14] is a distance measure between probability distributions and serves as a smoothed, symmetrized version of the Kullback-Leibler divergence. Properties of the JSD that differentiate itself from the KL divergence is that JSD is a symmetric distance measure and is bounded by 0 and 1 when using log (2). For two continuous probability distributions $P = \int_{-\infty}^{\infty} f(x)dx = 1$ and $Q = \int_{-\infty}^{\infty} g(x)dx = 1$, the JSD measure between P and Q is given by:

$$JSD(P||Q) = \frac{1}{2} D_{KL}(P||R) dx + \frac{1}{2} D_{KL}(Q||R) dx \qquad (1)$$

where $D_{KL}(P||Q)$ is the Kullback-Leibler distance between distributions $P$ and $Q$:

$$D_{KL}(P||Q) = \int_{-\infty}^{\infty} P(x) \log\left(\frac{P(x)}{Q(x)}\right) dx \qquad (2)$$

and $R$ is the mean distribution of $P$ and $Q$:

$$R(x) = \frac{1}{2}(P(x) + Q(x)) \qquad (3)$$

Expanded, the JSD can be represented as follows:

$$JSD(P||Q) = \frac{1}{2}\int_{-\infty}^{\infty} P(x)\, log\left(\frac{P(x)}{R(x)}\right) dx + \frac{1}{2}\int_{-\infty}^{\infty} Q(x)\, log\left(\frac{Q(x)}{R(x)}\right) dx \qquad (4)$$

Here, $JSD = 0$ if and only if $P$ and $Q$ are identical distributions, and $JSD(P||Q) = JSD(Q||P)$. The JSD determines the divergence between two probability distributions, $P$ and $Q$, where the divergence is near zero when a patient's given probability distributions are similar and increases as the distributions change. We use JSD across univariate distributions to detect anomalies in patient trajectory data. Given that we are calculating the JSD for a number of univariate features a comprehensive JSD distance at time $t$ will be obtained using the following equation.

$$D_{JS_{t_i,j}} = \frac{1}{M}\sum_j D_{JS_{t_i,j}} \qquad (5)$$

*Kernel Density Estimation*
In most literature, JSD is calculated as a measure of the difference between two explicitly specified probability distributions. Patient data is based on continuous, often sparse, empirical observations and does not provide explicit probability distributions. We utilized kernel density estimation (KDE) approaches to estimate the probability density of patient records in this research. As Parzen[15] and Rosenblatt[16] describe, KDE is a non-parametric technique for estimating the probability density function of a random variable using sample data. In practice, we estimate the JSD distance measure(s) based on the KDE results. Density estimates were calculated using Silverman's rule of thumb[17]. We selected this method in order to avoid making any distributional assumptions about the data since every patient's trajectory is distinct.

**Analysis**

Our proposed method for detecting infection employs a dynamic sliding window method and is based on comparing individual patient encounter characteristics to those of a reference group. These reference samples are drawn from a control cohort and represent individuals whose medical circumstances facilitated their admission into a hospital system's ICU and who had a SOFA score of two or greater at least 24 hours after admission, but whose trajectories improved sufficiently enough to avert infection.

*Data Preparation*
As previously stated, the data for this study is represented as a sparse matrix, necessitating extensive cleaning and preparation prior to analysis. Data was cleaned in three stages prior to the calculation of each probability distribution. We started by removing all features from the dataset that had more than 25% of their values missing. We then cleaned and imputed data at the patient and cohort levels, respectively.

*Patient Level*
At the patient level, if a patient's features contained at least 25% of their data, the mean of that feature was used to impute any missing values. For patient characteristics missing more than 75% of their data, those characteristics were excluded from patient-level imputations. As a result of the patient's analysis, information was dynamically excluded.

*Cohort Level*
At the cohort level, the dataset was duplicated and then divided into two subsets: the control cohort and the treatment cohort. Then, we identify features from both subsets for which more than 75% of their data was absent. These features were omitted from both the original and duplicated datasets. The means of the remaining features were utilized to impute any remaining missing data that had not been imputed at the patient level depending on which cohort a patient belonged to. Lastly, this new dataset was then used to calculate KDE of the control cohort's characteristics; it was not utilized for any other analyses.

*Critical Point Detection Algorithm using Jensen-Shannon Divergence*
Based on individual components, this algorithm seeks to identify significant deviations in probability distributions between two patient cohorts. Beginning with the combined cohort group, $G_c$, we first compute the KDE for each of the $M$ features. Next, determine the KDE for each feature of a single patient using the first two-period increments. The selection of a two-period time window was done without consideration of the bandwidth size of the KDE. Utilizing the KDE obtained from the cohort group and the KDE from the two-period increment obtained from the chosen patient, determine the JSD distance measure for each feature. Use Equation 5 to determine the total JSD score. For the chosen patient, increase the index by 1 and repeat the process $N$ times. The JSD algorithm's notation is described in Table 3, and the steps of the algorithm are described in Algorithm 1.

Table 3. JSD Algorithm Notations

| Variable | Definition |
|---|---|
| $M$ | number of features |
| $x_j^{G_c}$ | $j^{th}$ feature of the control group |
| $P_j$ | KDE of the $j^{th}$ feature of the control group |
| $N$ | number of patient hours in the ICU |
| $x_{i:i+1,j}^p$ | two-period window for a select patient's $j^{th}$ feature |
| $Q_{i+1,j}$ | KDE of the $j^{th}$ feature's two-period window |
| $D_{JS_{i+1,j}}^p$ | JSD distance measure for a select patient's $j^{th}$ feature |

**Algorithm 1.** Calculate JS divergence from sliding window

**for** $j = 1$ to $M$ **do**
  $P_j = KDE(x_j^{G_c})$
**end**
**for** $i = 1$ to $N$ **do**
  **for** $j = 1$ to $M$ **do**
    $Q_{i+1,j} = KDE(x_{i-1:i+1,j}^p)$
    $D_{JS_{i+1,j}}^p = JSD_{i,j}\left(P_j \parallel Q_{i+1,j}\right)$
  **end**
**end**

**Results**

Between 2016 and 2020, Grady Hospital admitted 17,845 patients to the intensive care unit. 893 patients met the aforementioned SOFA score and time horizon requirements in total. 749 patients were included in the control group, whereas 144 individuals were included in the treatment group. From the Grady database, we retrieved the following characteristics: demographic information (age, gender, race), ICU information (hourly vital signs, laboratory values, vasopressor values, and fluid administration values), Elixhauser comorbidity information, and discharge information.

In total, the JSD algorithm was used on 20 patients, with 10 patients coming from the control group and the other 10 patients coming from the treatment group. As the data provided was incremented hourly, we utilized a two-hour sliding window, beginning from each patient's ICU admission time, to determine each patient's respective probability distribution for time period, $t_i$. For each time period and feature, a single-sample JSD distance measure was computed. This calculation was carried out for each patient, yielding a total of 47 time periods, each representing an hour. A kernel density (KD) plot comparison of the aggregate comprehensive JSD distance measure for all control and treatment patients is shown in Figure 1. This plot empirically demonstrates the difference in the average deviation of a representative sample from each cohort when compared to the overall control group.

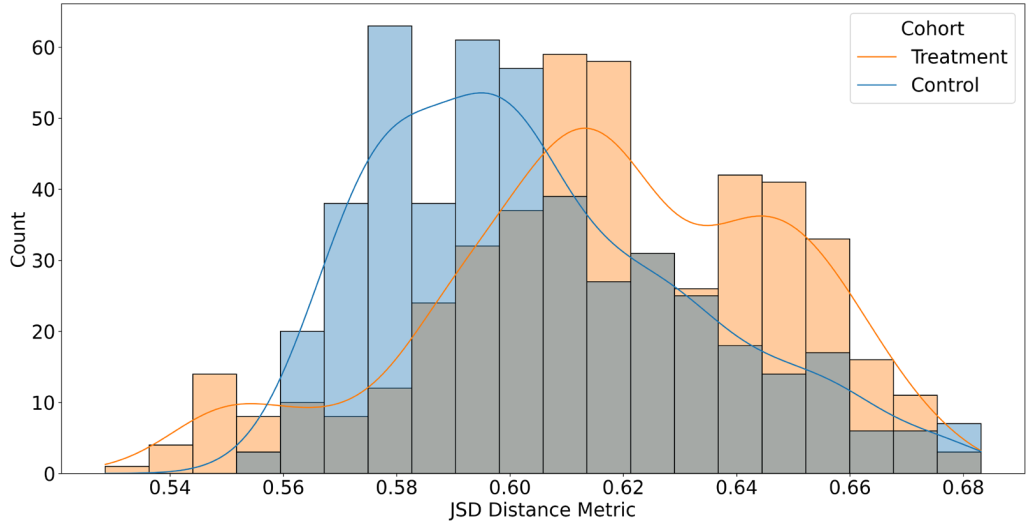

**Figure 1**. KD Plot of the average comprehensive JSD scores

The aggregated patients in the control and treatment cohorts are shown in Figure 2 along with a more detailed representation of single-sample JSD measures across a subset of physiologic factors, including calcium, phosphorous, hemoglobin, white blood cell count, body temperature, and sodium levels. By examining this subset of physiologic factors, the differences between the control and treatment groups are illustrated in greater detail.

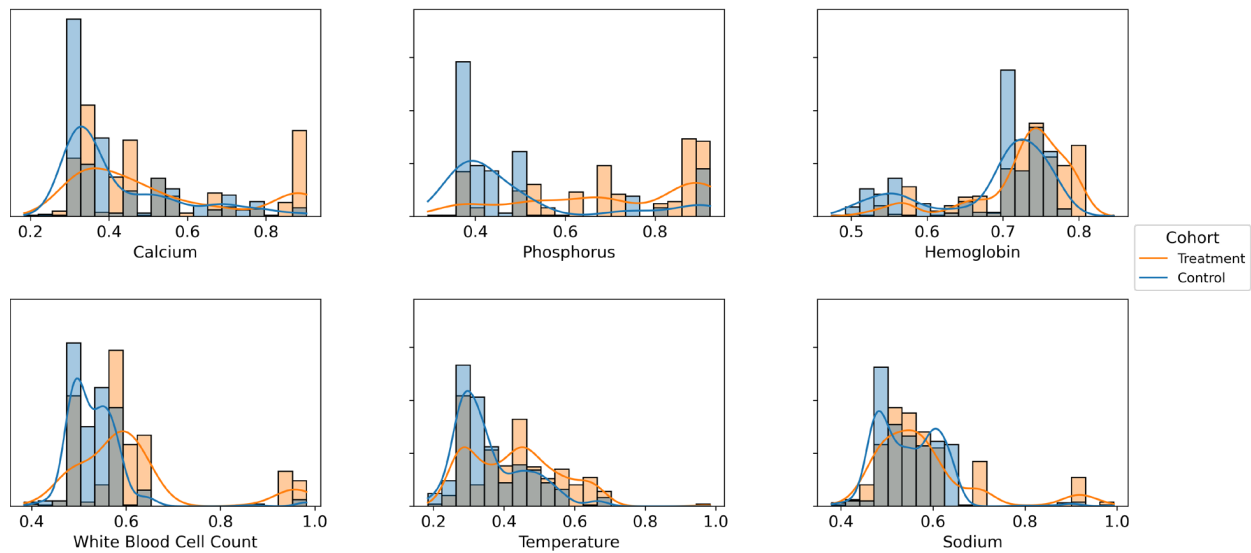

**Figure 2**. KD plot of physiologic feature subsets

The hourly quantitative measurement for each of the aforementioned physiologic factors is obtained by analyzing each cohort's combined JSD distance measure time-series data, as shown in Figure 3. On an hourly basis, these figures demonstrate the hourly separation and significant differences between the aggregated control and treatment cohorts.

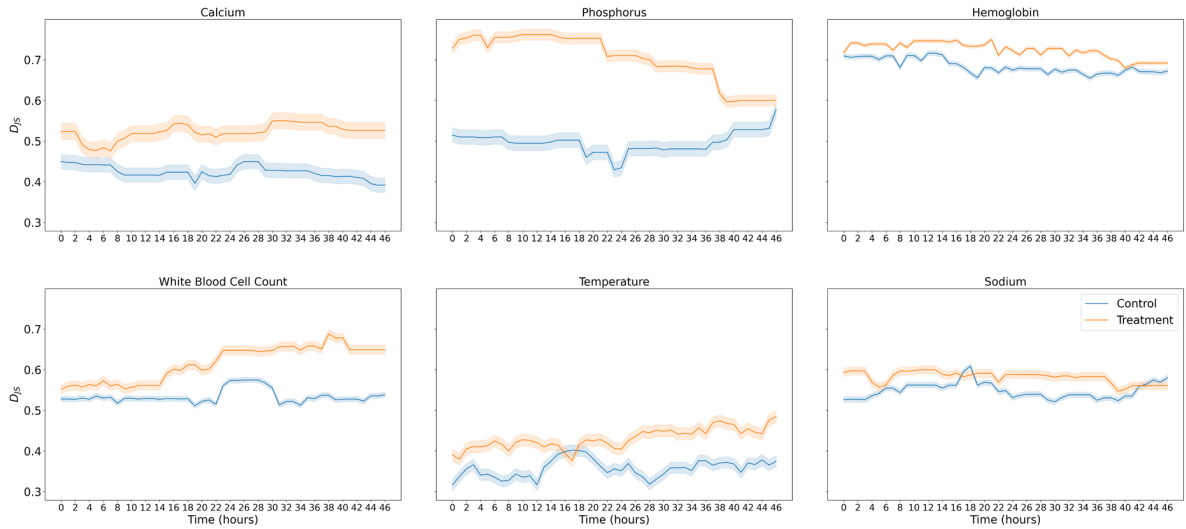

**Figure 3.** Time-series JSD score plots by cohort for six physiologic factors. JSD distance measurements represent the average of the JSD scores for the patients in each representative cohort. These averages are bounded by a 95% confidence interval.

Examining a randomly selected patient, patient k, from the treatment cohort, we can observe, as depicted in Figure 4, an increase in the comprehensive JSD distance measure approximately 7 hours following their admission to the ICU. A subset of the physiologic factors indicates an increase in deviation in the patient's calcium, hemoglobin, and sodium levels upon closer inspection. As represented in this figure, this visualization highlights how this method may enable a physician to quickly identify the individual physiologic abnormalities that lead to this prediction in an interpretable manner.

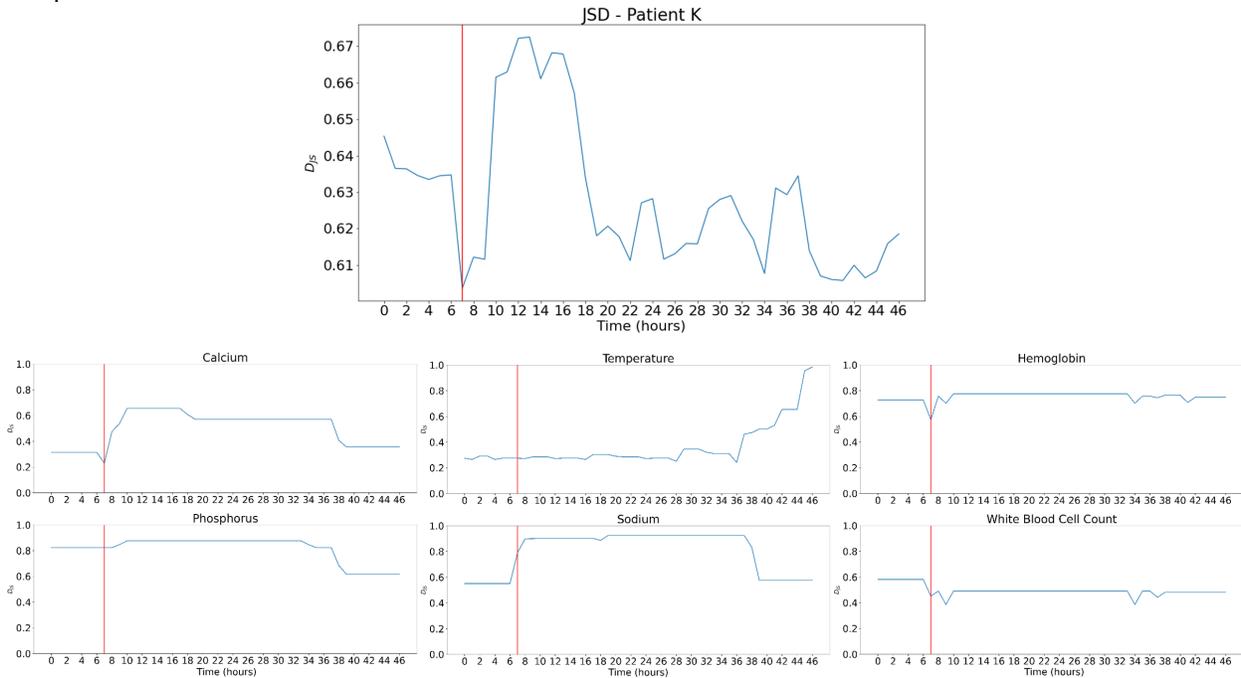

**Figure 4.** Example of a real-time JSD score in relation to likely contributing physiologic factors.

**Discussions**

This methodology requires that historical data from the reference group (control cohort) be continuously updated and maintained in a clinical workflow environment. As patients are admitted to the ICU, their vital signs and laboratory results are monitored in real-time and incorporated into the JSD algorithm using the two-period sliding window technique. An increase in a patient's JSD score denotes their departure from the status of patients in the same healthcare system who are never identified as having sepsis. Visual evaluation of the JSD algorithm's output provides clinicians with near-real-time indications of a patient's health deviation from the control group, as well as the physiologic factors that contribute most to the prediction. In addition, the two-period sliding window can be implemented with various time units, including seconds and minutes.

Monitoring dynamic changes in a patient's state is critical for detecting the development of an infection or the worsening of a disease. Several comparable studies have been conducted in this field, most notably by Yan, Li, Gao, Li, and Chen [18], who employed a single-sample JSD approach to develop a score index for detecting complicated illness early warning signals. Salem, Nait-Abdesselam, and Mehaoua[19] used JSD to identify anomalies in online network traffic to prevent flooding assaults in high-speed networks. This research employed a sequential analytic approach to detect abnormalities and a sliding window to provide a reference profile for comparison. Sunder and Shanmugam[20] used JSD to identify and avoid black hole attacks in wireless sensor networks used in healthcare. They applied their strategy to four distinct components and demonstrated an improvement in detection rate and a decrease in detection time for black hole assaults when compared to previous studies. JSD has been demonstrated to be an effective metric for estimating the divergence between a sample and reference distribution empirically. In this paper, we developed a JSD approach that quantitatively distinguishes between these two groups and provides a real-time measurement of divergence. Despite being well suited for this purpose, JSD can be data-sensitive due to the way it is constructed and lead to "division by zero" errors. We believe there is room for improvement in the ability to detect point-changes in the health status of ICU patients, despite the fact that this technique is interpretable and capable of handling both large and small datasets.

**Conclusion**

We proposed a sample-based Jensen-Shannon divergence distance measure based on single-sample time series data that can detect a patient's divergence from a reference group in real time and identify the causes most likely to contribute to their health deterioration. This method employs a non-parametric approach to determine distributions at multiple levels: reference groups, patient subsets, and individual patients, and provides clinicians with a patient-specific measure of health deterioration that they can not only interpret, but also trace back to its source.

**Acknowledgement**

The work of Yao Xie is partially supported by NSF DMS-2134037.